\documentclass[%
 reprint,
 amsmath,amssymb,
 aps,
]{revtex4-2}

\usepackage{graphicx}
\usepackage{dcolumn}
\usepackage{bm}
\usepackage{mathtools}
\usepackage{xcolor}
\usepackage{soul}
\sethlcolor{yellow}

\newcommand{\ist}[1]{\overset{\footnotesize(\ref{#1})}{=}}
\newcommand{\iist}[2]{\overset{^{(\ref{#1})}}{\underset{^{(\ref{#2})}}{=}}}

\begin{document}

\preprint{APS/123-QED}

\title{High-precision lattice determination of the interaction potential of an SU(2) solitonic dipole and comparison with perturbative QED}


\author{Manfried Faber}
 \email{faber@kph.tuwien.ac.at}
\author{Rudolf Golubich}%
 \email{rudolf.golubich@gmail.com}
 
\affiliation{%
Atominstitut, Technische Universit\"at Wien
}%




\date{\today}
\begin{abstract}
We determine the interaction potential of a solitonic dipole in the singlet state, modeled as an SU(2) field, using improved lattice simulations of two stationary solitons at varying separations. The potential is extracted from the energy of two-soliton configurations as a function of distance. At large separations, the interaction reproduces the classical Coulomb potential quantitatively up to an energy shift $\delta E_\infty\approx 9\;\text{keV}$ of the fitted asymptotic constant relative to $2m_ec_0^2$, assumed to be related to limited numerical precision on the lattice. At shorter distances, deviations from the Coloumb potential of point-like charges appear, that are in qualitative agreement with the asymptotic formula of perturbative Quantum Electrodynamics, reflecting the running of the fine-structure constant, with the inverse fine-structure constant ($\alpha^{-1} \approx 137$) reproduced.
\end{abstract}

\maketitle


\section{Introduction\protect}\label{sec:Introduction}
Previous work~\cite{Faber:1999ia,Faber:2022zwv} has introduced a model for topological particles through a 3+1-dimensional generalization of the 1+1-D Sine-Gordon model~\cite{remoissenet:1999wa}. These models demonstrate how localized field excitations, characterized by topological quantum numbers, possess relativistic properties characteristic of particles, such as stability, Lorentz contraction, and relativistic mass increase, thereby resolving the distinction between particles and fields.

The aim of this work is to compare the interaction of stationary topological solitons in the 3+1D model, using precise numerical calculations with as few approximations as possible, with the behavior of interacting electrons at rest and the predictions of QED in order to identify possible discrepancies between solitons and electrons. However, within the limits of numerical accuracy, the results show consistent agreement extending to the strength of the interaction and even to the behavior of the fine-structure constant~\cite{peskin}.

\subsection{Mathematical model}\label{sec:Mathematical}
To describe an SO(3)-valued field we use the fundamental representation of SU(2)
\begin{align}\label{eq:Field}
Q(x):&= e^{-\mathrm i\alpha(x)\, \vec{\boldsymbol\sigma}\cdot\vec n(x)} \notag\\
&= \underbrace{\cos\alpha(x)}_{q_0(x)}-\mathrm i\,\vec{\boldsymbol\sigma}
\cdot\underbrace{\vec n(x)\,\sin\alpha(x)}_{\vec q(x)}, |\vec n| = 1
\end{align}
with $x$ in Minkowski space-time, Pauli matrices $\vec{\boldsymbol\sigma}:=(\sigma_1, \sigma_2, \sigma_3)$ and arrows denoting components in the 3D Euclidean spaces, as in the $\mathfrak{su}(2)$-algebra. Solitons are described by the Lagrangian\,
\begin{equation}\label{eq:Lagrangian}
\mathcal{L}(x):=-\frac{\alpha_f\hbar c_0}{4\pi} 
\Biggl( \frac{1}{4}\,\vec R_{\mu\nu}(x)\cdot\vec R^{\mu\nu}(x)+\Lambda(x)\Biggr),
\end{equation}
with Sommerfeld's fine structure constant
\begin{equation}
\alpha_f:=\frac{e_0^2}{4\pi\varepsilon_0\,\hbar\,c_0},
\end{equation}
the potential
\begin{equation}\label{Potential}
\Lambda(x):=\frac{q_0^6(x)}{r_0^4}
\end{equation}
stabilizing the solitons core to a radius of $r_0$ and curvature tensor
\begin{equation}\label{GammaR}
\vec R_{\mu\nu}:=\vec\Gamma_\mu\times\vec\Gamma_\nu, \quad    
(\partial_\mu Q)Q^\dagger:=-\mathrm i\,\vec{\boldsymbol{\sigma}}\cdot\vec\Gamma_\mu,
\end{equation}
that is,
\begin{equation}\begin{aligned}
\vec\Gamma_\mu
&\iist{GammaR}{eq:Field}q_0\,\partial_\mu\vec q-(\partial_\mu q_0)\,\vec q
+\vec q\times\partial_\mu\vec q\\
&\ist{eq:Field}(\partial_\mu \alpha)\,\vec n
+\sin\alpha\cos\alpha\,\partial_\mu\vec n
+\sin^2\alpha\,\vec n\times\partial_\mu\vec n.
\end{aligned}\end{equation}

\subsection{Relation to physics}\label{sec:Relation}
As shown in Refs.\,\cite{Faber:1999ia,Faber:2022zwv}, the resulting nonlinear Euler–Lagrange equations admit four distinct stable single-soliton solutions
\begin{equation}\label{VierSol}
n_i(x)=\underbrace{Z}_{\pm 1}\frac{x^i}{|\vec x|},\quad\alpha(x)
=\begin{cases}\arctan\frac{|\vec x|}{r_0},\\\pi-\arctan\frac{|\vec x|}{r_0}.\end{cases}
\end{equation}
The rest energy of solitons is given by the analytical expression
\begin{equation}\label{eq:E0}
E_0=\frac{\alpha_f\,\hbar\,c_0}{r_0}\,\frac{\pi}{4}.
\end{equation}
From the solutions in  Eq.\,(\ref{VierSol}), it can be seen that the four topologically distinct field configurations have the same energy\,(\ref{eq:E0}). The proportionality
\begin{equation}\label{Feldst}
^*\vec F_{\mu\nu}:=-\frac{e_0}{4\pi\varepsilon_0c_0}\,\vec R_{\mu\nu}
\end{equation}
between the curvature tensor $\vec R_{\mu\nu}$ and the dual electromagnetic field strength tensor $^*\vec F_{\mu\nu}$ relates the mathematical quantities to the physical properties of the solitonic solutions. As a result of this relationship, static solitons are merely sources of electric fields, i.e., topologically quantized charges. Inside the solitons, the electromagnetic fields are non-Abelian, but become Abelian for distances much larger than the natural scale $r_0$ in Eq.\,(\ref{Potential}), i.e., under the influence of the potential.

The four topologically stable solutions differ in their charge quantum number $Z$ and in their coverings $\pm \frac{1}{2}$ of $\mathbb{S}^3$, whose absolute value defines the spin quantum number $s$ of the solitons. Two-soliton configurations are classified according to their total spin $S$, which is defined by the combined covering numbers of the individual solitons. This results in singlet states with total spin $S=0$ and triplet states with total spin $S=1$, as must follow from the triangular inequality for spin quantum numbers.

This similarity to electrons motivated comparisons with perturbative QED by adjusting the solitons rest mass to $ E_0 = m_e c_0^2=0.510\,998\,951\;\text{MeV}$. This is done by setting
\begin{equation}\label{natSkala}
r_0 = 2.213\,205\,16\;\text{fm},
\end{equation}
which according to Eq.\,(\ref{eq:E0}) is the classical electron radius multiplied by $\pi/4$.

Previous evaluations\,\cite{Wabnig2025} of the solitonic dipole potential suffered from limited precision due to numeric instabilities of the energy minimization procedures. In this work we present improved results for the solitonic dipole potential and a determination of a solitonic fine structure constant $\alpha_\mathrm{sol}$.

\section{Lattice formulation \protect}\label{sec:methodology}
\noindent
Exploiting the axial symmetry of the solitonic dipole, the field
configuration $Q(\varrho_r,z_k)$ is discretized on a static cylindrically symmetric lattice with spacing $a \leq \frac{r_0}{3}$,
\begin{equation}\begin{aligned}
q_0(\varrho_r,z_k)&= q_0(\vec x)\big|_{\vec x=(\varrho_r,0,z_k)},\\
q_\varrho(\varrho_r,z_k)&=\vec q(\vec x)\cdot\vec e_x\big|_{\vec x=(\varrho_r,0,z_k)},\\
q_z(\varrho_r,z_k) &=\vec q(\vec x)\cdot\vec e_z\big|_{\vec x=(\varrho_r,0,z_k)},
\end{aligned}\end{equation}
where the discrete coordinates are related to the lattice indices by
\(\varrho_r = r\,a\) for \(r = 0,\dots,n_\varrho-1\) and
\(z_k = (k-k_0)\,a\) for \(k = 1,\dots,n_z\) with
\(k_0 = (n_z+1)/2\).

The lattice size is chosen such that the distance of each soliton core to the nearest lattice boundary remains constant at $15 r_0 = 33.2\;\text{fm}$, while the separation between the soliton centers is varied. The lattice ranges from $45\times111$ (smallest) to $45\times475$ (largest).

It is initialized by embedding a single-soliton solution in each half of the lattice with the soliton centers separated by $d$ lattice spacings. The equilibrium field configuration is then determined by minimizing the discretized energy functional corresponding to the Lagrangian density $\mathcal{L}(x)$ of Eq.~(\ref{eq:Lagrangian}),
\begin{equation}\label{Eq:Egitter}
E[q_0,q_\varrho,q_z]=\underbrace{H_{\text{pot}}[q_0]}_{\propto\Lambda}
+\underbrace{H_{\text{curv}}[q_0,q_\varrho,q_z]}_{\propto\vec R_{\mu\nu}\cdot\vec R^{\mu\nu}}
+\underbrace{H_{\text{out}}[q_0]}_\text{exterior},
\end{equation}
using a nonlinear conjugate-gradient method with a bracketing and golden-section line search along the descent direction.

The potential contribution takes the form
\begin{equation}
H_{\text{pot}}[q_0]
\simeq \frac{\alpha_f\hbar c_0}{2r_0^4}\, a^2 \sum_{r=0}^{n_\varrho-1}
\sum_{k=1}^{n_z} r\, q_0^6(\varrho_r,z_k),
\end{equation}

The curvature term $H_{\text{curv}}$ is constructed from discrete
approximations to the derivatives $\partial_\varrho q_\mu$ and
$\partial_z q_\mu$ ($\mu \in \{0,\varrho,z\}$) entering the field
strength tensor $\mathbf{R}_{\mu\nu}$.
For interior lattice points we employ fourth–order accurate symmetric
five–point stencils in both directions,
\begin{align}
\partial_\varrho q_\mu(\varrho_r,z_k)
&\approx \frac{
    - q_\mu(\varrho_{r+2},z_k)
    + 8 q_\mu(\varrho_{r+1},z_k)
}{12\,a}
\nonumber\\
&\quad
  - \frac{
    8 q_\mu(\varrho_{r-1},z_k)
    - q_\mu(\varrho_{r-2},z_k)
}{12\,a},
\\[1.5ex]
\partial_z q_\mu(\varrho_r,z_k)
&\approx \frac{
    - q_\mu(\varrho_r,z_{k+2})
    + 8 q_\mu(\varrho_r,z_{k+1})
}{12\,a}
\nonumber\\
&\quad
  - \frac{
    8 q_\mu(\varrho_r,z_{k-1})
    - q_\mu(\varrho_r,z_{k-2})
}{12\,a},
\end{align}
while in the first and last lattice layers in $\varrho$ and $z$
we switch smoothly to three–point one–sided formulas of second order
in $a$,
\begin{align}
\partial_\varrho q_\mu(\varrho_0,z_k)
&\approx \frac{-3 q_\mu(\varrho_0,z_k)
 + 4 q_\mu(\varrho_1,z_k)}{2\,a}
\\
&\phantom{\approx}\mathrel{}
- \frac{q_\mu(\varrho_2,z_k)}{2\,a}, \\
\partial_\varrho q_\mu(\varrho_{n_\varrho-1},z_k)
&\approx \frac{q_\mu(\varrho_{n_\varrho-3},z_k)
 - 4 q_\mu(\varrho_{n_\varrho-2},z_k)}{2\,a}
\\
&\phantom{\approx}\mathrel{}
+ \frac{3 q_\mu(\varrho_{n_\varrho-1},z_k)}{2\,a},
\end{align}
and analogously for $\partial_z q_\mu$ at $z=z_1$ and $z=z_{n_z}$. Using these discrete derivatives, we construct the lattice analogue of $\vec R_{\varrho z}$ and hence of the curvature energy density
\begin{equation}
\mathcal{H}_{\text{curv}}(\varrho_r,z_k)
\propto\vec R_{\mu\nu}(\varrho_r,z_k)\cdot\vec R^{\mu\nu}(\varrho_r,z_k),
\end{equation}
which is then integrated over the cylindrical volume by summing over $r$ and $k$ with the appropriate Jacobian factor,
\begin{equation}
H_{\text{curv}}\simeq \frac{\alpha_f\hbar c_0}{4\pi}\, a^2
\sum_{r=0}^{n_\varrho-1}\sum_{k=1}^{n_z}r\,
\vec R_{\mu\nu}(\varrho_r,z_k)\cdot\vec R^{\mu\nu}(\varrho_r,z_k),
\end{equation}
which reproduces the continuum expression from Eq.~(\ref{eq:Lagrangian}) up to the discretization errors of the finite-difference scheme.

The contribution $H_{\text{out}}$ captures the energy outside the lattice by approximating it analytically via the Coloumb field of two charges at the location of the solitons cores at $z_\pm = \pm d/2$.

The radial and axial components of the electric field are
given by
\begin{align}
E_\varrho(\varrho,z)
&= \frac{e_0}{4\pi\varepsilon_0}
\left[
\frac{\varrho}{\bigl(\varrho^2+z_+^2\bigr)^{3/2}}
-\frac{\varrho}{\bigl(\varrho^2+z_-^2\bigr)^{3/2}}
\right],\\
E_z(\varrho,z)
&= \frac{e_0}{4\pi\varepsilon_0}
\left[
\frac{z_+}{\bigl(\varrho^2+z_+^2\bigr)^{3/2}}
-\frac{z_-}{\bigl(\varrho^2+z_-^2\bigr)^{3/2}}
\right].
\end{align}
The exterior electric-energy contribution is given by
\begin{equation}
  H_{\text{out}}
  = 2\pi \int_{\mathbb{R}^2\setminus\text{lattice}}
  \varrho\,\mathrm d\varrho\,\mathrm dz\; \frac{\varepsilon_0}{2}\bigl(E_\varrho^2(\varrho,z)+E_z^2(\varrho,z)\bigr),
  \end{equation} 
with $d\varrho$ from 0 to $R = (n_\varrho-\tfrac{1}{2})a$ and $dz$ from 0 to $Z = (n_z-\tfrac{1}{2})a$.

\section{Results\protect}
\begin{figure}[h]
\includegraphics[width=0.45\textwidth]{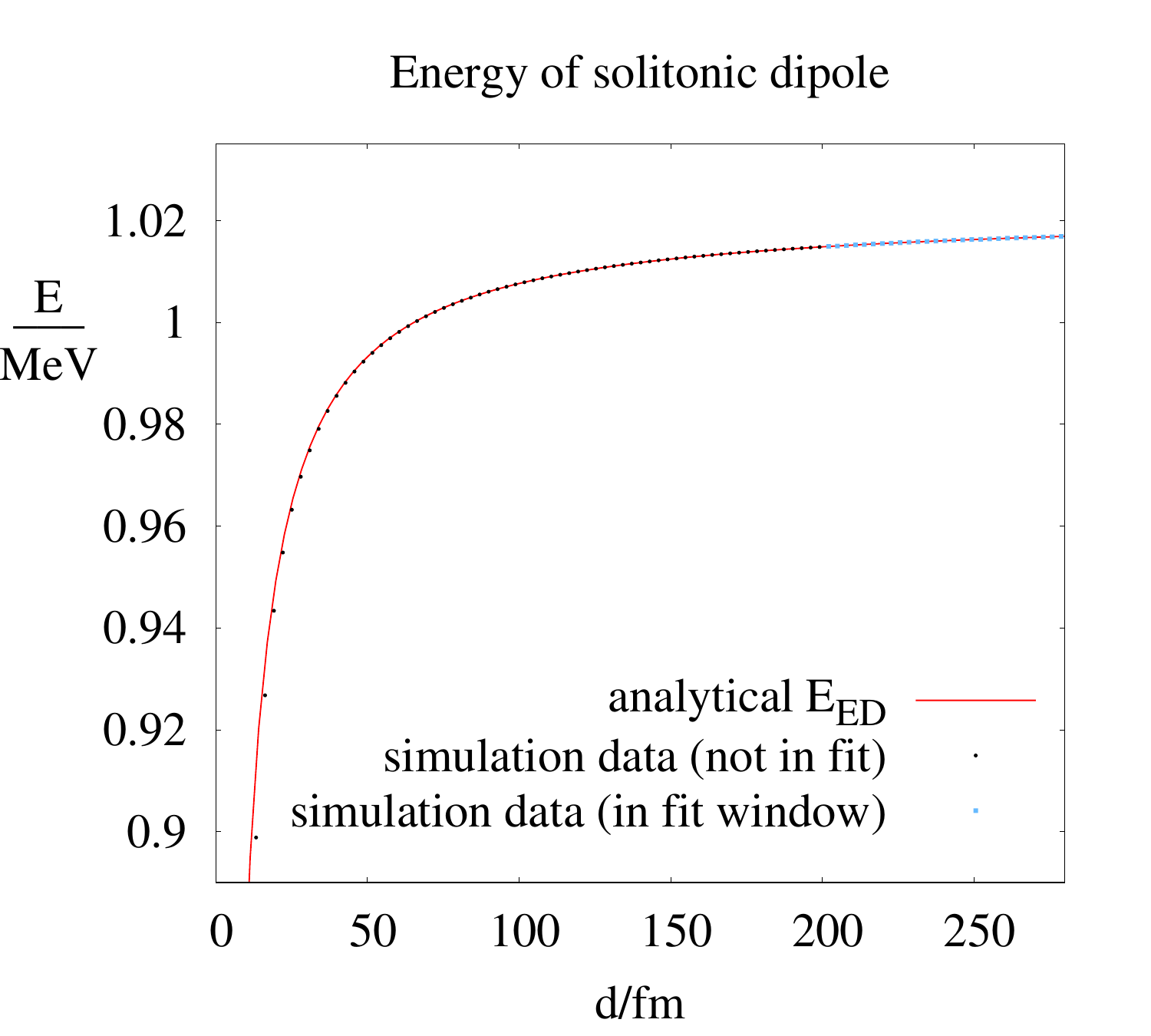}
\caption{The energies of solitonic dipoles with centers fixed at distance $d$ are compared to the energy $E_\mathrm{ED}(d)$ expected for a static dipole according to classical electrodynamics. Since discretization overestimates the energies of individual solitons, we shift the calculated energies by a corresponding constant $\delta E_\infty$, which is adjusted in the range $[200\text{--}280]\,\text{fm}$. Light blue squares: points included in the fit. Red line: fit to Eq.~(\ref{eq:Vsol}). Black circles: lattice points outside the fit window $[0\text{--}200]\,\text{fm}$. The markers are intentionally small to highlight deviations between analytical and numerical results at short distances.}
\label{fig:singene-coulomb}
\end{figure}
The energy $E(d)$ of solitonic dipoles at separation $d$ is determined according to Eq.\,(\ref{Eq:Egitter}), see Fig.\,\ref{fig:singene-coulomb}.
Due to the finite precision of the lattice simulation the extrapolated value at infinity overestimates the analytical minimum $2E_0$ of Eq.\.(\ref{eq:E0}) by a few keV, $E(\infty)=:2E_0+\delta E_\infty$ leading to the definitions
\begin{equation}\label{eq:Vsol}
E(d):=E[q_0,q_\varrho,q_z]-\delta E_\infty=2m_ec_0^2
-\frac{\alpha_{\text{sol}}\,\hbar c_0}{d}.
\end{equation}
$\delta E_\infty$ and $\alpha_{\text{sol}}$ are determinded for sufficiently large separations (typically $d\in[200,280]\;\text{fm}$) by a common fit leading to
\begin{equation}\label{eq:Fitwerte}
\delta E_\infty = 9.432(3)\,\text{keV},\quad
\alpha_{\mathrm{sol}}\hbar c_0=1.4387(8)\,\text{MeV\,fm}
\end{equation}
with uncertainties propagated from the nonlinear least-squares fit. In Fig.\,\ref{fig:singene-coulomb} the numerical results for $E(d)$ of Eq.\,(\ref{eq:Vsol}) are compared to the interaction of two point charges $\pm e_0$
at rest as described in electrodynamics by the the electron rest energy and the Coulomb potential
\begin{equation}
E_{\text{ED}}(d)=2m_ec_0^2-\frac{\alpha_f\hbar c_0}{d}.
\end{equation}
The good agreement of $\alpha_\mathrm{sol}\hbar c_0$ with $\frac{e_0^2}{4\pi\varepsilon_0}=1.43996$\,MeV\,fm is evidence that solitons interact like point charges for distances $d$ much larger then $r_0$ of Eq.\,(\ref{natSkala}).

\begin{figure}[h]
  \includegraphics[width=0.45\textwidth]{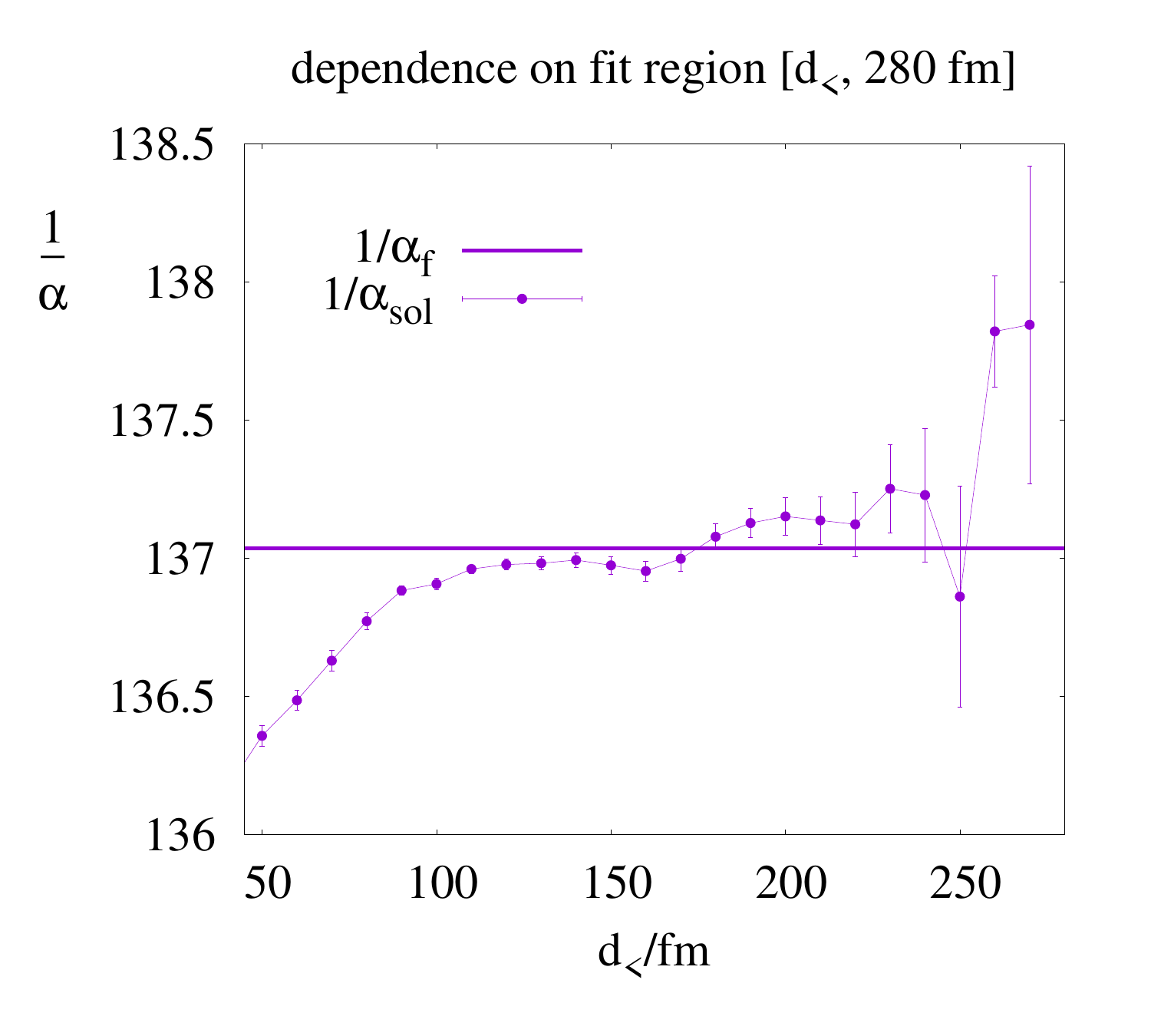}
  \caption{Dependence of the fit parameter $\alpha_\textrm{sol}$ of Eq.\,(\ref{eq:Vsol}) on the size of the fit region $[d_<, 280\,\textrm{fm}]$.}
  \label{fig:alle280}
\end{figure}

When adjusting the two parameters in Eq.\,(\ref{eq:Fitwerte}) in the fit region $[d_<, 280\,\textrm{fm}]$, one observes that $\delta E_\infty$ remains essentially unchanged over practically the entire range of reasonable $d_<$-values, whereas $\alpha_\mathrm{sol}\hbar c_0$ notes an interesting dependence. Since $1/\alpha_f$ is a famous number, we compare it in Fig.\,\ref{fig:alle280} with the dependence of $1/\alpha_\mathrm{sol}$ of Eq.\,(\ref{eq:Vsol}) on $d_<$. For $d_<$ less than 240\,fm, $1/\alpha_\mathrm{sol}$ begins to stabilize and then tends to decrease slightly, but from 80\,fm downward, a strong decrease in $1/\alpha_\mathrm{sol}$ becomes apparent which can no longer be attributed to numerical inaccuracies.

\begin{figure}[h]
  \includegraphics[width=0.45\textwidth]{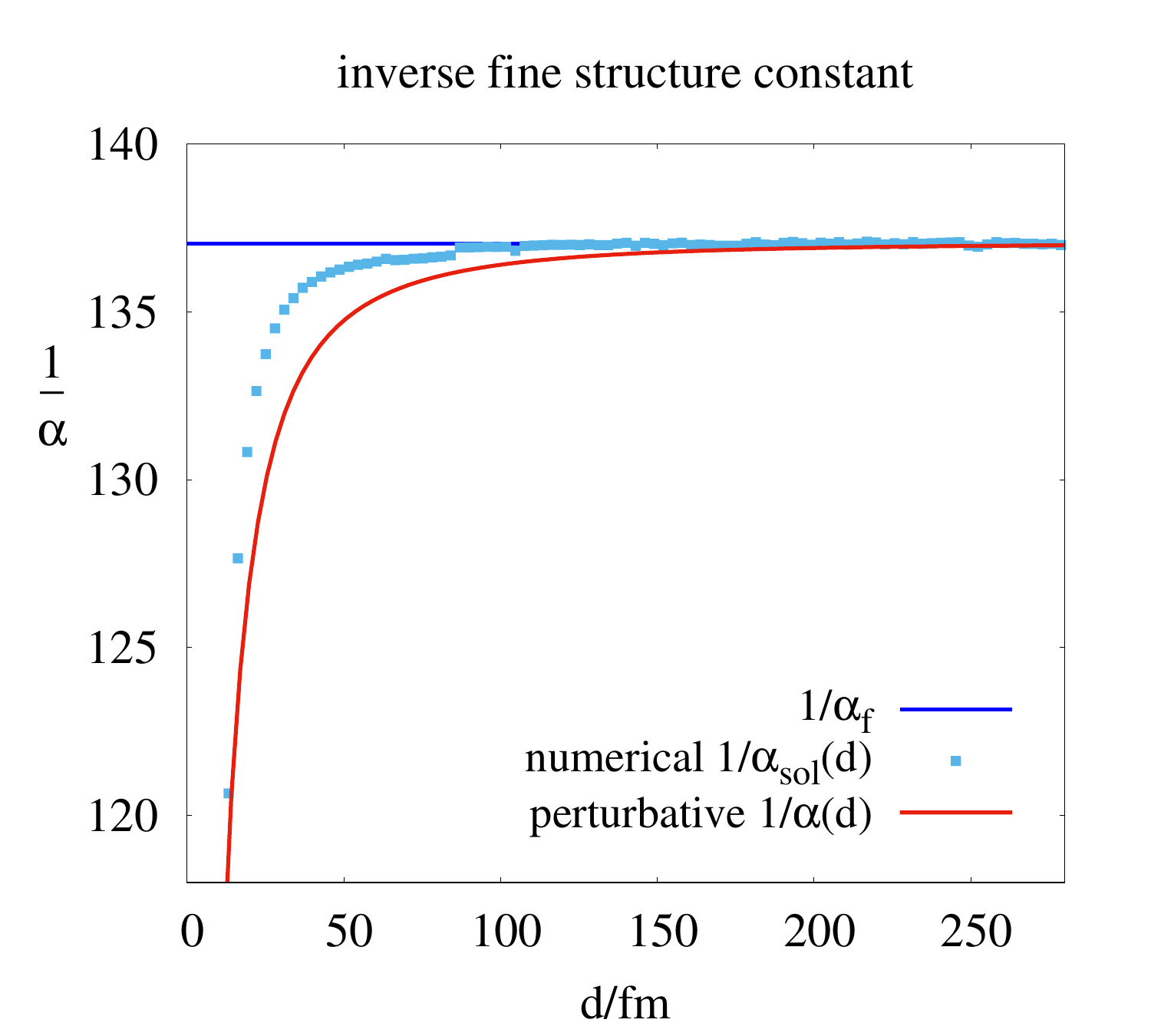}
  \caption{Comparison between the running coupling for the solitonic dipole and perturbative QED according to Eq.\,(\ref{eq:perturbativeQED}). For the numerical determination of $\alpha_\text{sol}(d)$ according to Eq. (\ref{eq:alphasold}), only the asymptotic dipole energy $\delta E_\infty$ is adjusted in the interval $[d_<, 280\,\textrm{fm}]$ with $d_<=200\,\textrm{fm}$, whereby the choice of $d_<$ does not significantly affect the position of the data points.}
  \label{fig:invalpha}
\end{figure}

At their center, due to their formulation by the non-Abelian scalar field $Q(x)$ of Eq.\,(\ref{eq:Field}), solitons are non-Abelian and have a finite radius. The extent of the non-Abelian field’s reach can be determined from the $d$-dependence of the fine-structure constant. We therefore introduce the quantity $\alpha_\text{sol}(d)$ into  Eq.\,(\ref{eq:Vsol}) and define
\begin{equation}\label{eq:alphasold}
\alpha_\text{sol}(d)=\frac{d}{\hbar c_0}\,[2m_ec_0^2-E(d)].
\end{equation}
In Fig.\,\ref{fig:invalpha} we compare $\alpha_\text{sol}(d)$ with the large distance approximation of perturbative QED~\cite[Eq. (7.95)]{peskin},
\begin{equation}\label{eq:perturbativeQED}
\alpha(d)=\alpha_f\Big\{1+\frac{\alpha_f}{\sqrt{2\pi}}
\frac{\mathrm e^{-2d/\Bar\lambda}}{(2d/\Bar\lambda)^{3/2}}\Big\}.
\end{equation}
with $\Bar\lambda=\hbar/(m_ec_0)$ the reduced Compton wavelength of the electron and the distance $d$ between the two solitons.

Figure~\ref{fig:invalpha} shows the running coupling and compares it to Eq.\.(\ref{eq:perturbativeQED}) of perturbative QED. The good agreement of $\alpha_\text{sol}(d)$ as $d\to\infty$ shows that the long-range Coulomb field is well reproduced in the numerical calculations. The strong $d$-dependence of $\alpha_\text{sol}(d)$ is, for sufficiently precise calculations, independent of the numerical approximation, as is also shown by the comparison with earlier, less accurate calculations; see Fig. 5 in \cite{Wabnig2025}. Its reason is the finite extent of the solitons. It is interesting how surprisingly well this dependence agrees with the predictions of perturbation theory.

\section{Discussion\protect}

The soliton model investigated in this work can be viewed as a further development of a dual formulation of Dirac’s magnetic monopoles~\cite{dirac:1931kp,dirac:1948um}. Dirac's magnetic monopoles have two singularities: the central singularity at the center, which is characteristic of all point charges, and the famous Dirac string, which leads to a quantization of the magnetic charge due to the requirement that it is invisible. Following a dual transformation, Dirac’s formulation thus explains the quantization of electric charge. In Wu and Yang's non-Abelian formulation of magnetic monopoles~\cite{wu1969solutions,Wu:1975vq}, they succeed in avoiding the Dirac string. As mentioned in~\cite{PhysRevD.13.3233} their ansatz is based on a suggestion by Hsu. Although they do not mention it, their gauge field can be derived from a hedgehog field consisting of unit vectors pointing in radial directions, and thus makes use of the property that an S$^2$ can be described without singularities using radial vectors.

Our soliton model~\cite{Faber:1999ia} is based on an additional step to eliminate the singularity at the origin. Instead of unit vectors, radial vectors are used, see the imaginary part of $Q$ in Eq.\,(\ref{eq:Field}), whose lengths approach zero at the origin and increase monotonically to a length of one as the distance from the center increases, thereby eliminating singularities. The topological solitons of this model are thus Dirac monopoles whose two singularities have been eliminated by the SU(2) and SO(3) formulations, respectively. The idea of defining electric and magnetic field strengths as area densities in the target space, see Eq.\,(\ref{GammaR}) and (\ref{Feldst}), was already formulated by Pisello in 1977 for an S$^2$ field~\cite{Pisello1977,Pisello1978}, whereas our work employs area densities on S$^3$.

The Skyrme model, which describes only short-range interactions, was proposed by Skyrme to account for nuclear forces and nucleons. It is also based on an SU(2) field~\cite{skyrme:1958vn,skyrme:1961vq}. Since Gauss’s law does not apply to the nuclear force, the concept of field strength cannot be meaningfully defined in the Skyrme model. In contrast to the Skyrmions, topological solitons reproduce the long range behavior of QED.

Extended magnetic monopoles --- regular field configurations of finite energy with long-range interactions, known as 't Hooft–Polyakov monopoles~\cite{Polyakov:1974ek,tHooft:1974kcl} --- were discovered in the Georgi–Glashow model~\cite{GeorgiGlashow}. Forces between these monopoles are influenced by the charged Higgs field. In the Prasad–Sommerfield limit, Manton finds that the force between monopoles of different charges follows an inverse quadratic law, but the forces between monopoles of equal charge vanish~\cite{Manton:1977}, which distinguishes them from topological solitons as well as charged particles in general.

Electrons, as described in conventional QED, are treated as point-like elementary particles without resolved internal structure. By contrast, the present topological solitons arise as extended, non-Abelian field configurations with finite core size. Nevertheless, within the precision of the observables investigated here, in particular the singlet-channel interaction potential and the extracted effective fine-structure constant, no measurable deviation has yet been identified that would permit a sharp phenomenological distinction between the solitonic configurations and electrons.

\section{Conclusion\protect}
We have presented high-precision lattice simulations of an SO(3) solitonic dipole. From the fitted Coulomb coefficient we extract an effective inverse solitonic fine-structure constant $\alpha_{\mathrm{sol}}^{-1}=137.1(1)$, within the numerical precision of the lattice consistent with the CODATA value~\cite{CODATA2022} $\alpha_f^{-1}=137.035999177(21)$. Furthermore, the effective coupling $\alpha_{\mathrm{sol}}(d)$ is derived from the lattice interaction energies and compared with the perturbative estimate Eq.~(\ref{eq:perturbativeQED}). $\alpha_{\mathrm{sol}}(d)$ reproduces the expected QED vacuum polarization at intermediate separations astonishingly well and recovers the asymptotic Coulomb behavior.

The solitonic dipole model reproduces key Coulomb and QED-scale features once the electron mass scale is related to the natural solitonic length scale through~$r_0$, supporting the interpretation of the SO(3) solitons as effective electron-like degrees of freedom in this framework.

Further investigations, including analysis of the triplet channel, are required to establish clear distinctions between solitonic and electronic configurations. Ongoing simulations aim to extract a solitonic analogue of the hyperfine splitting arising from the relative orientations of the spins of the electron and the positron.

\bibliography{singlet}

\end{document}